\documentstyle[12pt]{article}

 2

\def\pois#1#2{\{#1,#2\}}
\def\Tr{\hbox{{\rm Tr}}}
\def\Eq#1{{\begin{equation} #1 \end{equation}}}

\newcommand{\ba}{\begin{eqnarray}}
\newcommand{\ea}{\end{eqnarray}}
\begin{document}

\begin{flushright}
UAHEP 968\\
August 1996\\
\end{flushright}

\centerline{ \LARGE Bosonization and Current Algebra of Spinning Strings}
\vskip 2cm

\centerline{ {\sc  A. Stern}}

\vskip 1cm

\centerline{ Dept. of Physics and Astronomy, Univ. of Alabama,
Tuscaloosa, Al 35487, U.S.A.}

\vskip 2cm

\vspace*{5mm}

\normalsize
\centerline{\bf ABSTRACT}

We write down a general geometric action principle
for spinning strings in $d$-dimensional
   Minkowski space,
which is formulated without the use of Grassmann coordinates.
Instead, it is constructed in terms of the pull-back of a left invariant
Maurer-Cartan form on   the $d$-dimensional Poincar\'e  group to the
world sheet.
  The system contains some  interesting special cases.  Among them are
the Nambu string (as well as, null and tachyionic strings) where the
spin vanishes, and also the case of a  string with a spin current -
 but no momentum current.
  We find the general form for the Virasoro generators,
and show that they are first class constraints in the Hamiltonian
 formulation of the theory.  The current algebra
associated with the momentum and angular momentum densities are
shown, in general, to contain rather complicated anomaly terms which
obstruct quantization.  As expected, the anomalies vanish when one
specializes to the case of
the Nambu string, and there one simply recovers the algebra associated
with the Poincar\'e loop group.   We speculate that
there exist other cases
where the anomalies vanish, and that these cases give the bosonization
of the known pseudoclassical formulations of spinning strings.

\vskip 2cm
\vspace*{5mm}
\scrollmode

\newpage
\baselineskip=24pt
\section{Introduction}

The classical spin of  relativistic particles
can be described  using either classical or pseudoclasssical
variables.\cite{rbok}   The same result also
holds for the classical spin of relativistic
strings.  Of course, the pseudoclassical description of spinning strings
is well known.  Descriptions of spinning strings  in
terms of classical variables were examined recently in  ref. \cite{hs}.
     There we looked at the case
of $2+1$ space-time only, and the appropriate
 classical variables  took values in the $2+1$ Poincar\'e group $ISO(2,1)
$.    The string action
was constructed in terms of the pull-back of a left invariant
Maurer-Cartan form on $ISO(2,1)$ to the
world sheet.  Although,  it
has a particularly elegant form in $2+1$ dimensions due to the existence
of a nondegenerate scalar product on the Poincar\'e algebra
$\underline{ISO(2,1)}$ \cite{wit},  the action  can be generalized
to the case of an arbitrary number of
space-time dimensions, as well as to the case of membranes and p-branes.
In this article, we shall examine such generalizations.

As well as generalizing the system of ref. \cite{hs} to higher
dimensions, the approach described here generalizes that
developed by  Balachandran, Lizzi and Sparano\cite{baletal},
 which in turn gives a unifying description
 of the Nambu, null and tachyonic  strings.
We are able to recover the dynamics of ref. \cite{baletal} when we
specialize to the case of spinless strings.
\footnote {Spinning strings were
also considered in \cite{baletal} using a Wess-Zumino term.  Here we
shall show that there are more possibilities for introducing spin.}
   Our system also contains the case of pure spin,
 where there is a
nonvanishing spin current, but no momentum current.
 The existence of different cases is due to the different
choices available for the various constants present in the action.
These constants are the analogues of the mass and spin for the
relativistic  particle.

We write down the spinning string action in Sec 2.  Our criteria
 is that it be invariant
under Poincar\'e transformations, as well as, under diffeomorphisms.
The result is a straightforward generalization of the spinning
particle action described in \cite{rbok}.
 The resulting classical dynamics is obtained in Sec. 3
for four separate cases.  These cases
 include the  spinless string of ref.  \cite{baletal},
the case of pure spin, as well as the  most general  spinning string.

   As is usual, the Hamiltonian
description of the string system contains constraints.
  We proceed with the
constrained Hamiltonian dynamics in Sec. 4.  There we  write down the
general expression for the diffeomorphism
generators on a fixed time slice
of the string world sheet, and
we show that it satisfies the Virasoro algebra.  We then compute the
current algebra for the momentum and angular momentum densities for
the  four cases mentioned above.  For the case of the spinless string
  corresponding to ref.
 \cite{baletal}, we recover the algebra associated with the
  Poincar\'e loop group.
  On the other hand,
   for the case of pure spin, we get an extension of the Lorentz loop
group algebra.  The extension consists of
complicated anomalous terms which are obstructions to the quantization.
 Similar results are obtained for the
most general spinning string.  There we, instead, get an extension of
the Poincar\'e loop group algebra.

   If we demand that the above mentioned anomalous terms
 vanish for quantization, we  arrive at a
 set of equations for the constants defining the system.
These equations are quite complicated and we have not yet found their
 solutions.  We speculate that solutions exist and they correspond
to the bosonic formulation of known pseudoclassical descriptions
of spinning strings.   We note that the vanishing of the anomalous
terms of the classical Poincar\'e loop group algebra is a necessary
but not sufficient condition for a consistent quantization, as
new anomalies are expected to arise at the quantum level.  This indeed
is known to be
the case for  spinless strings.   We also note that a BRST approach
to the quantization of this system does not appear to be straightforward
due to the appearance of second class as well as first class constraints
in the Hamiltonian formalism.

It is straightforward to generalize our action for spinning strings to
 higher dimensional spinning objects, like membranes.
    We show how to do this in Sec. 5.

\section{ Classical Actions}

Before we write down the general expression for the
spinning string action, we give
 a discussion of our mathematical conventions and a brief review
of the classical description for the relativistic spinning
 particle.

\subsection{Mathematical Conventions}

As stated above, the target space for the spinning string will be taken
to be the Poincar\'e group.
We denote an element of the Poincar\'e group
 in $d+1$ space-time dimensions by $g=(\Lambda, x)$,
where  $\Lambda=\{{\Lambda^i}_j\}$ is a Lorentz matrix
and  $x=\{x^i\}$ a Lorentz vector, $i,j,..=0,1,2...n$.
$x$ will also serve to denote the Minkowski coordinates of the string.
We will only consider closed strings so $\Lambda $ and $x$ are
functions on $R^1 \times S^1$,  $R^1$ being associated with the time.

  Under the left action
of the Poincar\'e group by $h=(\theta, y)$, $g$
 transforms according to the semidirect product
rule: \Eq{(\Lambda, x)\rightarrow (\theta,y)\circ (\Lambda,x)
  =(\theta\Lambda, \theta x +y)\;.\label{lpt}}  Similarly,
under the right action of the Poincar\'e group by $h$,  we have that
\Eq{(\Lambda, x)\rightarrow (\Lambda,x)\circ   (\theta,y)
  =(\Lambda\theta, \Lambda y +x)\;.\label{rpt}}

Let $t_{ij}=-t_{ji}$ and $u_i$ denote a basis for
 the corresponding Poincar\'e algebra.
  For their commutation relations we have
   \ba
[t_{ij},t_{k\ell}] &=& \eta_{ik}t_{j\ell}  + \eta_{jk}t_{\ell i} +
\eta_{i\ell}t_{kj} +  \eta_{j\ell}t_{ik} \;,\cr
      [t_{ij},u_k] &=& \eta_{ik} u_j - \eta_{jk} u_i\;,  \cr
      [u_i,u_j] &=& 0\;,
 \ea
where $[\eta_{ij}] =$diag$(-1,1,1,...,1)$ is the Minkowski metric.

A left invariant
Maurer-Cartan form associated with the Poincar\'e group
can be written as follows:
\Eq{g^{-1}dg=W^{ij} t_{ij} +   V^iu_i\;,}
where the components $W^{ij} $ and $V^i$ are one forms given by
\Eq{W=\Lambda^{-1}d\Lambda\quad{\rm and} \quad V=\Lambda^{-1}dx \;.
\label{defVW}}
 It is  easy to check that $g^{-1}dg$ is
invariant  under left transformations (\ref{lpt}).
Under right transformations (\ref{rpt}),
  the Maurer-Cartan form transforms as  follows:
\Eq{g^{-1}dg\rightarrow g'^{-1}dg'=W^{ij} \;t'_{ij} + V^i\; u'_i\;} where
\ba
 t'_{ij}&=& (\theta t \theta^{-1} )_{ij} +
\frac12   (\theta u)_i  \;y_j  - \frac12   (\theta u)_j  \;y_i
  \;,\cr u'_i&=& (\theta u)_i \;. \ea
    The transformation from $t_{ij}$  and $u_{i}$ to $t'_{ij}$  and
     $ u'_{i}$
 by $h=(\theta, y)$ defines the adjoint
  action of the Poincar\'e group on the basis vectors.

We can now construct geometric actions which are
 invariant under Poincar\'e transformations.    For this,
 we let the action  depend only on
 the components $V$ and $W$ of $g^{-1}dg$.  It will then automatically
 be invariant under (left) Poincar\'e transformations (\ref{lpt}).
 With this prescription, we will arrive at a general description for
relativistic objects with spin.  To illustrate this we first review
 the case of relativistic particles.\cite{rbok}

\subsection{Spinning Particle Action}

The particle action is constructed in terms of the pull-back of
 $g^{-1}dg$ to the world line.
The most general geometric particle action which we  can write in this
way is
\Eq{S=S_{k}(\Lambda, x)=S_1+S_2  \;,\label{spa}} where
\ba S_1&=&\int k^{(1)}_{i}\; V^i \;, \\
S_2&=&-\frac12\int k^{(2)}_{ij}\;W^{ij}\;,    \ea   and
$k^{(1)}_{i}$ and $k^{(2)}_{ij}$ are constants, the latter being
antisymmetric in the indices $i$ and $j$.
In $d$ space-time dimensions there are then a total of
${1\over 2} d(d+1)$ constants.
These constants  determine the particle dynamics.  Actually for that
purpose, we only need to specify certain `orbits' of
$k^{(1)}_{i}$ and $k^{(2)}_{ij}$.
 These orbits are induced by
the action of the Poincar\'e group.  We define this action by the
following set of transformations $k\rightarrow k'$:
\ba k^{(1)}_{i}& \rightarrow & k^{'(1)}_{i}  =
[{}^\theta k^{(1)}]_{i}  \cr
k^{(2)}_{ij} &\rightarrow & k^{'(2)}_{ij}=
[{}^\theta k^{(2)}]_{ij}
-   [{}^\theta k^{(1)}]_{i}  \;y_j
+   [{}^\theta k^{(1)}]_{j}  \;y_i   \label{skt}   \;,\ea
where
\ba [{}^\theta k^{(1)}]_{i}&=&
{\theta_i}^r k^{(1)}_{r} \;,\cr
[{}^\theta k^{(2)}]_{ij}&=&
{\theta_i}^r{\theta_j}^s k^{(2)}_{rs} \;.\ea
$k^{(1)}_{i}$ and $k^{(2)}_{ij}$ thus transform under the Poincar\'e
group like momentum and angular momentum.
 The orbits can be classified by their
 invariants which are the usual ones for the Poincar\'e algebra.
In four space-time dimensions they are  $k^{(1)}_{i}k^{(1)i}$ and
$\omega_i\omega^i$, where $\omega^i
=\epsilon^{ijk\ell}k^{(1)}_{j}k^{(2)}_{k\ell}$.

  Using (\ref{skt}) it can be shown that
\Eq{S_{k'}(\Lambda, x) =S_{k}(\Lambda\theta,\Lambda y+ x)\;.\label{tspa}}
Thus, replacing $k^{(1)}_{i}$ and $k^{(2)}_{ij}$ in the action by
$k'^{(1)}_{i}$ and $k'^{(2)}_{ij}$
is equivalent to transforming the variables $\Lambda$ and $x$
by the right action (\ref{rpt}) of the Poincar\'e group.
  The equations of motion
obtained by varying $\Lambda'=\Lambda\theta$ and $x'=\Lambda y+x$ in
(\ref{tspa}) are identical to those resulting
 from variations of $\Lambda$
 and $x$ in (\ref{spa}).  Therefore (\ref{skt})
  define maps between equivalent dynamical systems.

The term $S_1$ in the action (\ref{spa}) by itself  describes a spinless
particle, while $S_2$ (which is
 a Wess-Zumino term for this system) gives rise to spin.
   This is easily seen from the
equations of motion, which we can obtain
by extremizing the action with respect to
 Poincar\'e transformations.    The equations of motion state
that there are two sets of constants of the motion.
From  infinitesimal translations, $\delta x^i=\epsilon^i$ and
$\delta {\Lambda^i}_j= 0$,
 we get the constants of
motion associated with the momentum,
\Eq{p_i=[{}^\Lambda k^{(1)}]_{i} \;.}
 From  infinitesimal Lorentz transformations, $\delta x^i=
{\rho^i}_j x^j$ , $\delta
{\Lambda^i}_j= {\rho^i}_k {\Lambda^k}_j$, for infinitesimal $\rho_{ij}=
-\rho_{ji}$,   we get constants of motion associated with
angular momentum,
\Eq{j_{ij} = x_i p_j-x_j p_i+ [{}^\Lambda k^{(2)}]_{ij} \;.}
$s_{ij}=[{}^\Lambda k^{(2)}]_{ij}$  then gives
 the spin contribution of the particle to the total angular momentum.

   As usual, $p_i$ and $j_{ij}$ are not all
independent.  For example, for a massive spinning particle they are
subject to two conditions:
\Eq{p^ip_i=k^{(1)}_{i}k^{(1)i} \qquad {\rm  and} \qquad
w^iw_i=\omega_i\omega^i\;,\label{copj}} where $w^i$ is the Pauli-Lubanski vector
 $w^i=\epsilon^{ijk\ell}p_{j}j_{k\ell}$. The quantum analogues of
  $p_i$ and $j_{ij}$  generate the Poincar\'e algebra in the
 quantum theory.  Their representations must be irreducible,
 the particular representation
being determined by the orbit of $k$ via conditions like
(\ref{copj}) on the quantum operators.

\subsection{Spinning String Action}

We now adapt a similar approach to the description of spinning strings.

A geometric action for strings can be expressed as a
 wedge products of the one   forms $V$ and $W$ defined in eq.
  (\ref{defVW}).  There are thus three possible terms:
\Eq{S=S_{K}(\Lambda, x)=S_1+S_2+S_3  \;,\label{ssa}} where
\ba S_1&=&{1\over 2}\int K^{(1)}_{ij}\; V^i \wedge V^j\;, \\
S_2&=&\int K^{(2)}_{ijk}\; V^ i \wedge W^{jk}\;,\\
S_3&=&{1\over 8} \int K^{(3)}_{ijk\ell}\; W^{ij} \wedge  W^{k\ell}\;.\ea
$K=(K^{(1)}_{ij},\;K^{(2)}_{ijk},\;K^{(3)}_{ijk\ell})$ denotes a set of
constants and they are the analogues of the constants
 $k^{(1)}_{i}$ and $k^{(2)}_{ij}$ appearing
in the particle action.  They satisfy the following symmetry properties:
\ba K^{(1)}_{ij}&=&-K^{(1)}_{ji}\;,\cr
 K^{(2)}_{ijk}&=&-K^{(2)}_{ikj}\;,\cr
K^{(3)}_{ijk\ell}&=&-K^{(3)}_{jik\ell}
=-K^{(3)}_{k\ell ij}=-K^{(3)}_{ij\ell k}\;.\label{sp}\ea
In $d$ space-time dimensions there are then a total of
${1\over 8} d(d-1)(d+1)(d+2)$ constants $K$.

The constants $K$ determine the string dynamics.  In analogy with the
particle description, we only need to specify the `orbits' on which
  $K$ lie.  These orbits are again induced by
the action of the Poincar\'e group.  We define this action by the
following set of transformations $K\rightarrow K'$:
\ba K^{(1)}_{ij}& \rightarrow & K^{'(1)}_{ij}  =
[{}^\theta K^{(1)}]_{ij}  \cr & &\cr
K^{(2)}_{ijk} &\rightarrow & K^{'(2)}_{ijk}=
[{}^\theta K^{(2)}]_{ijk}
+  {1\over 2}   [{}^\theta K^{(1)}]_{ij}  \;y_k
-  {1\over 2}   [{}^\theta K^{(1)}]_{ik}  \;y_j      \;,\cr & &\cr
K^{(3)}_{ijk\ell} &\rightarrow & K^{'(3)}_{ijk\ell} =
[{}^\theta K^{(3)}]_{ijk\ell}  \;
+\; 2 [{}^\theta K^{(2)}]_{ik\ell} \;y_j +
   2 [{}^\theta K^{(2)}]_{\ell i j} \;y_k -
   2 [{}^\theta K^{(2)}]_{kij} \;y_\ell -
   2 [{}^\theta K^{(2)}]_{jk\ell}\;y_i \cr
& &\qquad \quad  +
[{}^\theta K^{(1)}]_{ik} \;y_j y_\ell-
[{}^\theta K^{(1)}]_{i\ell} \;y_j y_k -
[{}^\theta K^{(1)}]_{jk} \;y_i y_\ell +
[{}^\theta K^{(1)}]_{j\ell} \;y_i y_k \;, \label{sskt}  \ea
where
\ba [{}^\theta K^{(1)}]_{ij}&=&
{\theta_i}^r{\theta_j}^s K^{(1)}_{rs} \;,\cr
[{}^\theta K^{(2)}]_{ijk}&=&
{\theta_i}^r{\theta_j}^s{\theta_k}^t K^{(2)}_{rst} \;,\cr
[{}^\theta K^{(3)}]_{ijk\ell} &=&
{\theta_i}^r{\theta_j}^s{\theta_k}^t {\theta_\ell}^u K^{(3)}_{rstu}\;.\ea
   Using  these definitions it can be shown that
\Eq{S_{K'}(\Lambda, x) =S_{K}(\Lambda\theta,\Lambda y+ x)\;,\label{tssa}}
in analogy to (\ref{tspa}).  Thus, replacing $K$ in the action by $K'$
is equivalent to transforming the variables $\Lambda$ and $x$
by the right action (\ref{rpt}) of the Poincar\'e group.
  The equations of motion
obtained by varying $\Lambda'=\Lambda\theta$ and $x'=\Lambda y+x$ in
(\ref{tssa}) are identical to those resulting from variations of $\Lambda$
 and $x$ in (\ref{ssa}).  Therefore (\ref{sskt}) define maps
  between equivalent dynamical systems.

   The orbits induced by the action (\ref{sskt})  of the Poincar\'e
group  on $K$ can be labeled by their invariants.
One simple quadratic invariant is of course $$\Tr K^{(1)^2}\;.$$
In general, the expression for the invariant depends on
the number of space-time dimensions.  For the example of three space-time
dimensions, we found the following additional quadratic invariant:
\cite{hs}
$$\epsilon^{ijk}\epsilon^{\ell mn} ( K^{(2)}_{\ell jk}
 K^{(2)}_{imn} - {1\over 4}  K^{(3)}_{ijk\ell} K^{(1)}_{mn} )\;.$$
 In four space-time dimensions, one instead has
the quadratic invariant:\cite{baletal}
 $$\epsilon^{ijk\ell} K^{(1)}_{ij} K^{(1)}_{k\ell}\;.$$

\section{Equations of Motion}

We next examine the classical string dynamics following from the actions
i) $S_1$,  ii) $S_2$, iii) $S_3$ and iv) $S=S_1+S_2+S_3  $.

i)  The action $S_1$, which can be expressed by
\Eq{S_1={1\over 2}  L^{(1)}_{ij}\;dx^i \wedge dx^j\;,}
 was discussed in ref. \cite{baletal} and for certain orbits of $K$
 is known to be equivalent
to the Nambu action.   Here we define
$ L^{(1)}_{ij}=   [{}^\Lambda K^{(1)}]_{ij}=
{\Lambda_i}^r{\Lambda_j}^s K^{(1)}_{rs} \;.$
The  standard form of the Nambu action is obtained upon eliminating
${\Lambda^i}_j$ (which in this case play the role of auxiliary
variables) from  $S_1$.  For  other orbits of $K$, the action $S_1$ can
  yield either tachyonic or null strings\cite{null}.

  For all choices of $K^{(1)}_{ij}$ the action $S_1$ alone
    describes a spinless
string.  This will be evident from the form of the conserved momenta
and angular momenta.  These conserved currents are found
by extremizing the action with respect to
 Poincar\'e transformations.
From  infinitesimal translations $\delta x^i =\epsilon^i$ and
$\delta {\Lambda^i}_j= 0$,  we get the equations of
motion corresponding to momentum current conservation,
\Eq{\partial_\alpha P_{(1)i}^\alpha=0\;,\quad\;
P_{(1)i}^\alpha=\epsilon^{\alpha\beta}\;L^{(1)}_{ij}\;
\partial_\beta x^j \;,\label{pone}}
 where $\alpha,\beta,..$ denote world sheet  indices
 and $\epsilon^{01}= - \epsilon^{10}=  1$.
 From  infinitesimal Lorentz transformations, $\delta x^i=
{\rho^i}_j x^j$ , $\delta
{\Lambda^i}_j= {\rho^i}_k {\Lambda^k}_j$, for infinitesimal $\rho_{ij}=
-\rho_{ji}$,   we get angular momentum current conservation,
\Eq{\partial_\alpha J_{(1)ij}^\alpha=0\;,\quad\;
J_{(1)ij}^\alpha = x_i P_{(1)j}^\alpha  -  x_j P_{(1)i}^\alpha\;.  }
Consequently, the angular momentum current consists of only an orbital
term, and  therefore the string defined by the action $S_1$ is spinless.

ii)  A spin contribution to the angular momentum current is present for
strings described by the action $S_2$, which can also be written as
\Eq{S_2=L^{(2)}_{ijk}
 \; dx^i \wedge (d\Lambda \Lambda^{-1})^{jk}\;,}
where $L^{(2)}_{ijk}= [{}^\Lambda K^{(2)}]_{ijk}
={\Lambda_i}^r{\Lambda_j}^s{\Lambda_k}^t K^{(2)}_{rst}  $.
Now from infinitesimal translations $\delta x^i=\epsilon^i$ and
$\delta {\Lambda^i}_j= 0$, we get the equations of
motion \Eq{\partial_\alpha P_{(2)i}^\alpha=0\;,\quad\;
P_{(2)i}^\alpha=\epsilon^{\alpha\beta}  \;
L^{(2)}_{ijk} \;(\partial_\beta\Lambda\Lambda^{-1})^{jk} \;.\label{ptwo}}
  From infinitesimal Lorentz transformations, $\delta x^i=
{\rho^i}_j x^j$ , $\delta{\Lambda^i}_j={\rho^i}_k {\Lambda^k}_j$,
 we get that
\Eq{\partial_\alpha J_{(2)ij}^\alpha=0\;,\quad\;
J_{(2)ij}^\alpha  =  x_i P_{(2)j}^\alpha  -  x_j P_{(2)i}^\alpha
+S_{(2)ij}^\alpha\;. }
$S_{(2)ij}^\alpha$ denotes the
 spin contribution to the angular momentum  current.  It is given by
\Eq{S_{(2)ij}^\alpha  =2\epsilon^{\alpha\beta}\;  L^{(2)}_{kij} \;
\partial_\beta x^k      \;.\label{stwo}}

iii)  We next consider the case of strings described by the action $S_3$,
which can be written \Eq{S_3=\frac18 L^{(3)}_{ijk\ell}
 \;(d\Lambda \Lambda^{-1})^{ij} \wedge (d\Lambda \Lambda^{-1})^{k\ell}\;,
}  where $L^{(3)}_{ijk\ell} = [{}^\Lambda K^{(3)}]_{ijk\ell}
={\Lambda_i}^r{\Lambda_j}^s{\Lambda_k}^t {\Lambda_\ell}^u K^{(3)}_{rstu}
\;.$  Now the angular
momentum current consists solely of a spin term.  This is evident
because the momentum current vanishes, $P_{(3)i}^\alpha=0$, and hence
so does the orbital angular momentum.  Here the
spin current has the following form:
\Eq{J_{(3)ij}^\alpha  =S_{(3)ij}^\alpha
 =-{1\over 2}\epsilon^{\alpha\beta} \;L^{(3)}_{ijk\ell}
\; (\partial_\beta\Lambda\Lambda^{-1})^{k\ell}  \;,\label{sthre}}
which is seen by extremizing $S_3$ with respect to variations
$\delta{\Lambda^i}_j= {\rho^i}_k {\Lambda^k}_j$.
In $2+1$ space-time dimensions, $S_3$ was seen to be the integral
of an exact two form, and as a result there $J_{(3)ij}^\alpha$
was identically conserved.\cite{hs}

iv)  For the most general action (\ref{ssa})
 consisting of all three terms $
S_1,\;S_2$ and $S_3\;,$ the conserved momentum and angular momentum
currents are given by the sum of the individual currents:
\ba P_{i}^\alpha &=&P_{(1)i}^\alpha   +P_{(2)i}^\alpha
=\epsilon^{\alpha\beta}\biggl(L^{(1)}_{ij}
\;\partial_\beta x^j    + L^{(2)}_{ijk} \;
(\partial_\beta\Lambda\Lambda^{-1})^{jk}\biggr) \;,\\
J_{ij}^\alpha &=&J_{(1)ij}^\alpha +J_{(2)ij}^\alpha
 +J_{(3)ij}^\alpha\;.\ea
The angular momentum  current can be expressed as a sum of an orbital
part $x_i P_{j}^\alpha  -  x_j P_{i}^\alpha $ and a
spin current.  The latter is given by
 \Eq{S_{ij}^\alpha =S_{(2)ij}^\alpha +S_{(3)ij}^\alpha
 =
\epsilon^{\alpha\beta}\biggl( 2 \;L^{(2)}_{kij} \;
\partial_\beta x^k
-{1\over 2} \;L^{(3)}_{ijk\ell}
\; (\partial_\beta\Lambda\Lambda^{-1})^{k\ell}   \biggr)
 \;.}

We note that the tensors $L^{(1)}_{ij}  $, $L^{(2)}_{ijk}$ and
$L^{(3)}_{ijk\ell} $ which appear in the string equations of motion
satisfy the same symmetry properties as
the constant tensors $K^{(1)}_{ij}  $, $K^{(2)}_{ijk}$ and
$K^{(3)}_{ijk\ell} $, i.e., we can replace $K$ by $L$ in (\ref{sp}).

\section{Constraint Analysis and Current Algebra }

We next proceed with the Hamiltonian formulation of the system.
Constraints are present in the Hamiltonian description
 since all terms in the action (\ref{ssa}) are
 first order in world sheet time ($\tau$-)
derivatives.
  Furthermore, the constraints can be either first or second class.
  As a result of this,
the analysis of the current algebra is quite involved.

   Below we shall handle
the cases i)-iv) separately.  Before doing so, however,
we shall guess the
expression for the diffeomorphism generators on a fixed time slice
of the string world sheet, and
show that it satisfies the Virasoro algebra.

We first  introduce the momentum variables $\pi_i$ and $\Sigma_{ij}$,
 which along
with $x^i$ and $\Lambda$ span the  phase space.
 The variables $\pi_i$ are defined to be
canonically conjugate to $x^i$, while $\Sigma_{ij}$
generate left transformations on $\Lambda$.  This can be expressed
in terms of the following equal (world sheet) time Poisson brackets: \ba
\pois{x_i(\sigma)}{\pi_j(\sigma')}&=&
\eta_{ij}\; \delta(\sigma-\sigma')\;,\cr
\pois{\Lambda(\sigma)}{\Sigma_{ij}(\sigma')}&=&[t_{ij}\Lambda](\sigma)
\; \delta(\sigma-\sigma')\;,\cr
\pois{\Sigma_{ij}(\sigma)}{\Sigma_{k\ell}(\sigma')}&=&[
 \eta_{ik}\Sigma_{j\ell}  + \eta_{jk}\Sigma_{\ell i} +
\eta_{i\ell}\Sigma_{kj} +  \eta_{j\ell}\Sigma_{ik}](\sigma)\;
 \delta(\sigma-\sigma')\;,\label{canpb}\ea
where we write the phase space variables on a world sheet time slice, the
phase space variable being periodic functions of  the spatial
coordinate $\sigma$ of the world sheet.
All other Poisson brackets between the phase space variables are zero.
Below we shall utilize the following matrix representation for $t_{ij}$:
\Eq{(t_{ij})_{k\ell}= -\eta_{ik} \eta_{j\ell } + \eta_{i\ell}\eta_{jk}\;.}

As mentioned above, the constraints can be classified as both
 first and  second class.   In this regard, we know
from the reparametization symmetry of the action,  that there
exist some combinations of the constraints
which are  first class.   Those which generate diffeomorphisms
along a $\tau=constant$ surface should satisfy the Virasoro algebra.
It is easy to construct such generators.  They are:
\Eq{
{\cal L}[f] =\int d\sigma \;f(\sigma) \biggl(\pi_i  \partial_\sigma x^i -
\frac12 \Sigma_{ij}( \partial_\sigma \Lambda
\Lambda^{-1})^{ij}\biggr)\;,\label{difgen} }
  where $f$ are  periodic functions
on a fixed $\tau$ slice.  From (\ref{canpb}), it can be verified
 that ${\cal L}[f]$ indeed do  satisfy the Virasoro algebra,
\Eq{\pois{{\cal L}[f]} {{\cal L}[f']} ={\cal L}[f\partial_\sigma f'
-f'\partial_\sigma f ]\;.\label{va}}

  In the subsections that  follow,
we shall  show that
${\cal L}[f] $ are  first class constraints.
For this we shall need  the specific
form for the action in order to obtain
the explicit expressions for the constraints, which look different
in the four different cases mentioned earlier.
With this in mind we now specialize  to the cases i)-iv).

\subsection{Case i)}
As stated before, case i) describes a spinless string.
The constraints on the momentum variables are:   \ba
\psi_i &=& \pi_i - L^{(1)}_{ij}\partial_\sigma x^j
  \approx 0\;,\cr
\phi_{ij}&=& \Sigma_{ij}
 \approx 0\;.\label{con1}\ea
From the first constraint and (\ref{pone}),  $\pi_i $ is thus
identified with the time component of the momentum current
$P^0_{(1)i} $.
In addition to  $\pi_i$,
we can define the   variables \Eq{{\cal J}_{ij} =
 x_i \pi_j - x_j \pi_i + \Sigma_{ij} \;,\label{jdef}}
which are  weakly equal
(i.e., up to a linear combination of constraints)
to the time component of the angular momentum current $J^0_{(1)ij}$.

To compute their Poisson brackets, we find it convenient to
write the constraints as distributions: \ba
\Psi[\lambda]&=&\int d\sigma \lambda^i(\sigma)\psi_i(\sigma)\approx 0
\;,\cr      \Phi[\rho]&=&\int d\sigma \rho^{ij}(\sigma) \phi
_{ij}(\sigma)\approx 0\;,\label{condis}\ea
where $\lambda^i$ and $\rho^{ij}=-\rho^{ji}$
 are periodic functions of the spatial coordinate $\sigma$.
From the Poisson brackets (\ref{canpb}), we obtain the following
algebra of the constraints:  \ba
\pois{\Psi[\lambda]}{\Psi[\lambda']}&=& \int d\sigma (\partial_\sigma
\lambda^j \lambda'^i -\partial_\sigma \lambda'^j \lambda^i) L^{(1)}_{ij}
\;,\cr
\pois{\Psi[\lambda]}{\Phi[\rho]}&\approx & 2\int d\sigma \rho^{ij}
 (\lambda_i \pi_j + \lambda^k\partial_\sigma x_i L^{(1)}_{kj}) \; \;,\cr
\pois{\Phi[\rho]}{\Phi[\rho']}&\approx &    0\;.\ea
Note that the last two relations only hold weakly.

From the reparametization symmetry of the action, we know that there
exist linear combinations of
$\Psi[\lambda]$ and $\Phi[\rho]$ which are  first class constraints.
Let  \Eq{\Theta [\hat \lambda,\hat\rho]=\Psi[\hat \lambda]+
\Phi[\hat\rho]\label{fcc}} be a general first class constraint,
   where the Lagrange multipliers
 $\hat \lambda$  and $\hat \rho$ denote the functions
 $\hat\lambda^i(\sigma)$ and
 $\hat\rho^{ij}(\sigma)=-\hat\rho^{ji}(\sigma)$, respectively.
   They are solutions to the following equations:
\ba{{\hat \rho}_i} ^{\;j} \pi_j &\approx &\frac12\hat\lambda^j\partial
_\sigma L^{(1)}_{ji} +\hat\rho^{jk}\partial_\sigma x_k
L^{(1)}_{ij}\;,\cr
\hat\lambda_i \pi_j -\hat\lambda_j \pi_i
&   \approx &\hat\lambda^k ( L^{(1)}_{jk}   \partial_\sigma x _i -
 L^{(1)}_{ik}   \partial_\sigma x _j  )     \;,\label{hat1}\ea  which
follow from demanding that $\Theta$ is first class.
Since these equations are linear in $\hat\lambda$ and $\hat\rho$,
it follows that if $\hat\lambda$ and $\hat\rho$ are  solutions to
 (\ref{hat1}),  then $f\hat\lambda$ and $f\hat\rho$ are also solutions to
 (\ref{hat1}), where $f$ is an arbitrary function on the world sheet.
 The generators of diffeomorphisms of the world sheet then
  have the general form:  $\Theta[f\hat \lambda,f\hat\rho]$.

One solution to equations (\ref{hat1}) is $\hat \lambda = \partial_\sigma x$
and $\hat \rho=-\frac12 \partial_\sigma \Lambda \Lambda^{-1}$.  The
resulting first class constraint (corresponding to $f=1$),
\Eq{H_\sigma  =\Theta[\hat \lambda,\hat\rho]
=\Psi[\partial_\sigma x ]+\Phi\biggl[-\frac12 \partial_\sigma \Lambda
\Lambda^{-1}\biggr]=\int d\sigma \biggl(\pi_i  \partial_\sigma x^i -
\frac12 \Sigma_{ij}( \partial_\sigma \Lambda
\Lambda^{-1})^{ij}\biggr) \;,\label{hsig}}
  generates translations in $\sigma$:
\ba \partial_\sigma x^i(\sigma)& =&\pois{ x^i(\sigma)}{H_\sigma} \;,\cr
\partial_\sigma {\Lambda^i}_j
(\sigma)& =&\pois{ {\Lambda^i}_j(\sigma)}{H_\sigma} \;.\label{sigt}\ea
When $f\ne 1$ we get  the generators of diffeomorphisms
\Eq{\Theta[f\hat \lambda,f\hat\rho]  = {\cal L}[f] \;,}
on a fixed $\tau$ slice of the world sheet.
 $ {\cal L}[f] $ were given in (\ref{difgen}) and
   satisfy the Virasoro algebra (\ref{va}).  We have therefore shown that
the Virasoro generators are first class constraints.

For the case where $\Theta[\hat \lambda,\hat\rho]=H_\tau$ is
 the generator of $\tau$ translations, we need that
 $\hat\lambda_i$  is a time-like vector.
By computing the Hamilton equations of motion for $\pi_i(\sigma)$
and $ {\cal J}_{ij}(\sigma)$, we can recover
 the current conservation law for momentum and angular momentum.
This follows because the Poisson brackets of
$\pi_i(\sigma)$ and $ {\cal J}_{ij}(\sigma)$ with the constraints
are spatial derivatives,  \ba
\pois{\Psi[\lambda]}{\pi_i(\sigma)}&=& \partial_\sigma (\lambda^k
L^{(1)}_{ki}) \;,\cr
\pois{\Phi[\rho]} {\pi_i(\sigma)}&=&0\;,\cr
 \pois{\Psi[\lambda]} {{\cal J}_{ij}(\sigma)  }
&\approx &  \partial_\sigma\biggl(\lambda^k (x_i L^{(1)}_{kj} -
x_j L^{(1)}_{ki} )\biggr) \;, \cr
\pois{\Phi[\rho]}{{\cal J}_{ij} (\sigma)}&\approx &0 \;.\ea
The Hamilton equations are then:
\ba\partial_\tau \pi_i(\sigma) & =& \pois{ \pi_i(\sigma)}{H_\tau} =
 -\partial_\sigma(\hat\lambda^k L^{(1)}_{ki} )\;,\cr
\partial_\tau {\cal J}_{ij}(\sigma)& = &
 \pois{{\cal J}_{ij}(\sigma) }{H_\tau}\approx
 -\partial_\sigma\biggl(2\hat\lambda^{k}(x_i L^{(1)}_{kj} -
 x_j L^{(1)}_{ki})    \biggr)\;.\ea
  Here we get the identification of
 $\hat\lambda^k L^{(1)}_{ki} $ with the space component of the
momentum current  $ P^1_{(1)i}$ defined in (\ref{pone}).
These equations once again show that the case i) string is spinless.

The generators of the Poincar\'e symmetry are the  charges
\Eq{\int d\sigma \; \pi_i \quad{\rm and }\quad
\int d\sigma \; {\cal J}_{ij} \;.\label{char}} They
have zero Poisson brackets with the constraints.
This again follows because the Poisson brackets of
$\pi_i(\sigma)$ and $ {\cal J}_{ij}(\sigma)$ with the constraints
are spatial derivatives.  The charges
(\ref{char}) are thus first class variables as well as
Dirac variables.

  It remains to construct the Dirac variables associated
with current densities $\pi_i(\sigma)$ and ${\cal J}_{ij}(\sigma) $.
 For this we first define
\Eq{\tilde \pi[\gamma]=\int d\sigma\; \gamma^i(\sigma)
\pi_i(\sigma)+\Phi[E_{(\gamma)}]\;.}
$E_{(\gamma)} $ denote the functions
$E_{(\gamma)}^{ij}(\sigma)=-E_{(\gamma)}^{ji}(\sigma)$.
  $\tilde \pi[\gamma]$
have (weakly) zero Poisson brackets with $\Phi[\rho]$.
For them to have zero Poisson brackets with $\Psi[\lambda]$, we
need that the functions $E_{(\gamma)}^{ij}(\sigma)$ satisfy:
\Eq{-\frac12 \partial_\sigma \gamma^i L^{(1)}_{ki}+ E_{(\gamma)kj}\pi^j
+E_{(\gamma)}^{ij} L^{(1)}_{kj}  \partial_\sigma x_i \approx 0\;.}
Similarly, we can define the variables
 \Eq{\tilde {\cal J}[\xi] =  \int d \sigma\;\xi^{ij}(\sigma)
 {\cal J}_{ij}(\sigma)  +\Phi[G_{(\xi)}]\;.}
  $G_{(\xi)}$ denote the functions $G_{(\xi)}^{ij}(\sigma)=
  -G_{(\xi)}^{ji}(\sigma)$.
Like $\tilde \pi[\gamma]$,
 they have (weakly) zero Poisson brackets with $\Phi[\rho]$.
For them to have zero Poisson brackets with $\Psi[\lambda]$, we
need that  $G_{(\gamma)}^{ij}(\sigma)$ satisfy:
\Eq{ G_{(\xi)ki}\pi^i +G_{(\xi)}^{ij} \partial_\sigma x_i L^{(1)}_{kj} -
\partial_\sigma\xi^{ij} x_iL_{kj}^{(1)} \approx 0 \;.\label{gone}    }
$\tilde\pi[\gamma]$ and $\tilde{\cal J}[\xi] $ are then
Dirac variables.  Since they have zero Poisson brackets with first, as
well as second class constraints, they are gauge invariant.

We next compute the Poisson bracket (or current) algebra for the
momenta and angular momenta.  For this purpose
it turns out, in this case, not to be necessary to solve (\ref{gone})
 for $G_{(\xi)ki}$.     The current algebra is simply the algebra of
the  Poincar\'e loop group: \ba
\pois{\tilde \pi[\gamma]}{\tilde \pi[\gamma']}& =& 0 \;,\cr
\pois {\tilde {\cal J}[\xi]} {\tilde \pi[\gamma]}
&\approx & -2\tilde\pi [\xi \gamma]   \;,\cr
\pois {\tilde {\cal J}[\xi]}   {\tilde {\cal J}[\xi']}& \approx &
-4\tilde {\cal J} [\xi\xi '] \;.\ea

$\tilde {\cal J}[\xi]$ and $\tilde \pi[\gamma]$ are gauge invariant
coordinates which label the reduced phase space.
Whether or not they form a complete set of variables is not evident.
On the other hand, we note that for spinless strings
$\tilde {\cal J}[\xi]$ and $\tilde \pi[\gamma]$ are not independent
on the reduced phase space, since from (\ref{con1}) and (\ref{jdef}), are $
{\cal J}_{ij}$
and $\pi_k$ are subject to:        \Eq{
 {\cal J}_{ij}\pi_k  + {\cal J}_{jk}\pi_i   + {\cal J}_{ki}\pi_j
 \approx 0\;.}

\subsection{Case ii)}
This case is of interest because it yields a nontrivial momentum and
angular momentum current, the latter containing a spin contribution
$S^\alpha_{(2)ij}$.
Here we shall  derive a current algebra which is an extension of
the  Poincar\'e loop group.

In this case the constraints on the momentum variables are:
\ba \psi_i& = &\pi_i -
 L^{(2)}_{ijk}(\partial_\sigma \Lambda \Lambda^{-1})^{jk}
  \approx 0\;,\cr
\phi_{ij}&= &\Sigma_{ij} - 2L^{(2)}_{kij}\partial_\sigma x^k
 \approx 0\;.\ea
From the first constraint and (\ref{ptwo}),  $\pi_i $ is thus
identified with the time component of the momentum current
$P^0_{(2)i} $, while from the second constraint and (\ref{stwo})
$\Sigma_{ij}$ is  identified with the time component of the
 spin current $S^0_{(2)ij}$.
In addition to  $\pi_i$ and $\Sigma_{ij}$,
we can once again define the   variables ${\cal J}_{ij}$,
 as was done in (\ref{jdef}), which are  weakly equal
to the angular momentum densities $J^0_{(2)ij}$.

For the algebra of the constraints we now get:
\ba\pois{\Psi[\lambda]}{\Psi[\lambda']}&=& 0\;,\cr
\pois{\Psi[\lambda]}{\Phi[\rho]}&\approx &2\int d\sigma \rho^{ij}
 (\lambda_i \pi_j - \lambda^k\partial_\sigma  L^{(2)}_{kij}) \; \;,\cr
\pois{\Phi[\rho]}{\Phi[\rho']}&\approx    &
4\int d\sigma \biggl((\rho^{ij}{\rho '}^{k\ell} -
{\rho '}^{ij}\rho  ^{k\ell})L^{(2)}_{\ell ij} \partial_\sigma x_k
-{{\rho'}^i}_k \rho  ^{kj} \Sigma_{ ij} \biggr)\;,\label{aoctwo}\ea
where $\Psi[\lambda]$ and $\Phi[\rho]$ are once again the
 distributions defined in (\ref{condis}).
As with  case i), from the reparametization symmetry of the action, we
know that there exist linear combinations of
$\Psi[\lambda]$ and $\Phi[\rho]$ which are  first class constraints.
First class constraints may also arise due to additional
symmetries associated with some particular choices for $K^{(2)}$.
For  $\Theta [\hat \lambda,\hat\rho]$ as defined in (\ref{fcc})
to be a  first class constraint,
   we need that the Lagrange multipliers
 $\hat \lambda$  and $\hat \rho$ now satisfy:
\ba& &{{\hat \rho}_i} ^{\;j} \pi_j \approx \hat\rho^{jk}\partial_\sigma
L^{(2)}_{ijk}\;,\cr    & & \cr             & &
 \frac12(\hat\lambda_i \pi_j -\hat\lambda_j \pi_i)
 - {{\hat\rho}_i}^{\;k}\Sigma_{kj}    + {{\hat\rho}_j}^{\;k}\Sigma_{ki}
 \cr  & & \quad \approx \hat\lambda^k \partial_\sigma L^{(2)}_{kij}
-\hat\rho^{k\ell} (L^{(2)}_{jk\ell} \partial_\sigma x _i
- L^{(2)}_{ik\ell}\partial_\sigma x_j
+2 L^{(2)}_{kij} \partial_\sigma x _\ell)  \;.\label{hat2}\ea
Since, like (\ref{hat1}), these equations are linear in
 $\hat\lambda$ and $\hat\rho$,
it again follows that if  $\hat\lambda$ and $\hat\rho$ are
 also solutions to  (\ref{hat2}), then so are
$f\hat\lambda$ and $f\hat\rho$
 , where $f$ is an arbitrary function on the world sheet.
 The generators of diffeomorphisms of the world sheet will again
  have the general form:  $\Theta[f\hat \lambda,f\hat\rho]$.
 $\hat \lambda = \partial_\sigma x$
and $\hat \rho=-\frac12 \partial_\sigma \Lambda \Lambda^{-1}$ are once
again solutions to the above conditions, with
the resulting first class constraint (corresponding to $f=1$) being equal to
  $H_\sigma$ in (\ref{hsig}).  As before, $H_\sigma$
  generates translations on a $\tau=constant$ slice of the world
sheet    (\ref{sigt}).
When $f\ne 1$ we recover the generators of diffeomorphisms
${\cal L}[f] =\Theta[f\hat \lambda,f\hat\rho] $
on a fixed $\tau$ slice of the world sheet, which satisfy
 the Virasoro algebra (\ref{va}).  Thus we have again shown that
the Virasoro generators are first class constraints.

For the case where $\Theta[\hat \lambda,\hat\rho]=H_\tau$ is
 the generator of $\tau$ translations, we again need that
 $\hat\lambda_i$  is a time-like vector.
To obtain the  Hamilton equations of motion for $\pi_i(\sigma)$
and $ {\cal J}_{ij}(\sigma)$, we can use         \ba
\pois{\Psi[\lambda]}{\pi_i(\sigma)}&=& 0 \;,\cr
\pois{\Phi[\rho]} {\pi_i(\sigma)}&=&2\;
\partial_\sigma (L^{(2)}_{ijk}\rho^{jk})\;,\cr
  \pois{\Psi[\lambda]} {{\cal J}_{ij}(\sigma)  }
&\approx & -2\partial_\sigma(\lambda^k
L^{(2)}_{kij} ) \;, \cr
\pois{\Phi[\rho]}{{\cal J}_{ij} (\sigma)}&\approx &-2 \; \partial_\sigma
\biggl(\rho^{k\ell}(L^{(2)}_{ik\ell}
 x_j - L^{(2)}_{jk\ell} x_i ) \biggr) \;.\label{pwctwo}\ea
Thus as in the previous case, the Poisson brackets of
$\pi_i(\sigma)$ and $ {\cal J}_{ij}(\sigma)$ with the constraints
are spatial derivatives.
The current conservation law for momentum and angular momentum then
follows:                                    \ba
\partial_\tau \pi_i(\sigma) & =& \pois{ \pi_i(\sigma)}{H_\tau} =
 -\partial_\sigma(2L_{ijk} \hat\rho^{jk})\;,\cr
\partial_\tau {\cal J}_{ij}(\sigma) &=&
 \pois{{\cal J}_{ij}(\sigma) }{H_\tau}\approx
 -\partial_\sigma\biggr(2\hat\rho^{k\ell}(x_i L_{jk\ell}
 -x_j L_{ik\ell}) -2\hat\lambda^k L_{kij}\biggr)\;.\ea
 Here we get the identification of
 $2L^{(2)}_{ijk} \hat\rho^{jk}$ with the space component of the
 momentum current $ P^1_{(2)i}$, defined in (\ref{ptwo}),
  and  $-2\hat\lambda^k L_{kij}$ with
the space component of the spin current $ S^1_{(2)ij}$,
defined in (\ref{stwo}).

As before, the charges (\ref{char}) are the  generators
 of the Poincar\'e symmetry.
From (\ref{pwctwo}),  they
have zero Poisson brackets with the constraints,
and  therefore they are
Dirac variables.  We can construct the Dirac variables associated
with current densities $\pi_i(\sigma)$ and ${\cal J}_{ij}(\sigma) $,
in a similar manner to what was done previously.
 We  define                                           \Eq{
\tilde \pi[\gamma]=\int d\sigma\; \gamma^i(\sigma)
\pi_i(\sigma)+\Psi[F_{(\gamma)}]\;.}
$F_{(\gamma)} $ denote the functions
$F_{(\gamma)}^j(\sigma)$.  From the first equations in (\ref{aoctwo})
 and (\ref{pwctwo}), $\tilde \pi[\gamma]$
have zero Poisson brackets with $\Psi[\lambda]$.
For them to have zero Poisson brackets with $\Phi[\rho]$, we
need that the functions $F_{(\gamma)}^j(\sigma)$ satisfy:\Eq{
\partial_\sigma \gamma^i L^{(2)}_{ijk} +\frac12 (F_{(\gamma)j}\pi_k
-F_{(\gamma)k}\pi_j) -F_{(\gamma)}^i\partial_\sigma L^{(2)}
_{ijk}\approx 0\;.}
We next define the variables
 \Eq{\tilde {\cal J}[\xi] =  \int d \sigma\;\xi^{ij}(\sigma)
 {\cal J}_{ij}(\sigma) +\Psi[H_{(\xi)}] +\Phi[G_{(\xi)}]\;.}
  $H_{(\xi)}$ and  $G_{(\xi)}$ denote the functions
$H_{(\xi)}^i(\sigma)$ and $G_{(\xi)}^{ij}(\sigma)=
-G_{(\xi)}^{ji}(\sigma)$, respectively.
For  $\tilde {\cal J}[\xi]$
 to have zero Poisson brackets with all of the constraints, we need
that    $H_{(\xi)}$ and  $G_{(\xi)}$ satisfy:               \ba
& &\partial_\sigma\xi^{ij} L_{kij}^{(2)} + G_{(\xi)ki}\pi^i -G_{(\xi)}^{ij}
\partial_\sigma L^{(2)}_{kij} \approx 0\;,\cr     & &\cr
& &2\partial_\sigma\xi^{ij}L^{(2)}_{ik\ell}x_j
-\frac12 (H_{(\xi)k}\pi_\ell  - H_{(\xi)\ell}\pi_k)
+H_{(\xi)}^i\partial_\sigma L^{(2)}_{ik\ell}\cr
& & -G_{(\xi)}^{ij}(L^{(2)}_{\ell ij}\partial_\sigma x_k
-L^{(2)}_{k ij}\partial_\sigma x_\ell-2
L_{jk\ell}^{(2)}\partial_\sigma x_i)-G_{(\xi)ik}{\Sigma^i}_\ell
+G_{(\xi)i\ell}{\Sigma^i}_k\approx 0 \;. \label{GHeq}   \ea
Due to the existence of the first class constraints $\Theta$,
$H_{(\xi)}$ and  $G_{(\xi)}$ are not uniquely defined by these equations.

The current algebra generated by the momentum
and angular momentum  densities can be written in terms of the
functions $F_{(\gamma)} $,
  $H_{(\xi)}$ and  $G_{(\xi)}$.  Unlike in the spinless case, we now find
a rather complicated result:                   \ba
\pois{\tilde \pi[\gamma]}{\tilde \pi[\gamma']}& =& 0 \;,\cr
\pois {\tilde {\cal J}[\xi]} {\tilde \pi[\gamma]}          &
\approx &-2\tilde\pi [\xi \gamma] - 2 \tilde \pi [G_{(\xi)}F_{(\gamma)}]
 - 2 \int d \sigma\; G_{(\xi)}^{ij}F_{(\gamma)}^k \partial_\sigma
 L^{(2)}_{kij}\;,\cr
\pois {\tilde {\cal J}[\xi]}   {\tilde {\cal J}[\xi']} &\approx &
-4\tilde {\cal J} [\xi\xi ']+2\tilde\pi [G_{(\xi ')}  H_{(\xi)}] -
2\tilde\pi [G_{(\xi )}  H_{(\xi')}]   \cr & &
+2\int d \sigma\;\biggl(
 H_{(\xi)}^i G_{(\xi ')}^{jk} \partial_\sigma L^{(2)}_{ijk}  +2
 G_{(\xi )}^{ij} G_{(\xi ' )}^{k\ell} L^{(2)}_{jk\ell}
  \partial_\sigma x_{i}  +
 G_{(\xi )}^{ij}{ G_{(\xi ' )j}}^{k} \Sigma_{ ki}
   - (\xi \rightleftharpoons \xi')\biggr)      \cr & &
                          \label{etpatwo}                     \ea
We have used equations (\ref{GHeq}) to write
the last Poisson bracket in a manifestly  antisymmetric way.

The first terms on the right hand sides of equations  (\ref{etpatwo})
  correspond to the algebra associated
 with the  Poincar\'e loop group, while the
  remaining terms represent a complicated
  extension of that algebra.  If we make the assumption that the momentum
and angular momentum currents of    relativistic
strings must satisfy the Poincar\'e loop group algebra in the quantum
theory, then the above remaining terms are anomalous and they are obstructions
to the quantization of the theory.  Requiring that all the
anomalous terms vanish in (\ref{etpatwo})  yields a total of
${1\over 8} d(d-1)(d^2 +3d-2)$ conditions on the $ {1\over 2} d^2(d-1)$
functions $L^{(2)}_{ijk}$, and then on the $ {1\over 2} d^2(d-1)$
constants $K^{(2)}_{ijk}$.  The system looks overdetermined, however
that is not entirely clear because, as stated earlier,
the functions $H_{(\xi)}$ and  $G_{(\xi)}$ are not uniquely defined by
(\ref{GHeq}).  Even if the system is overdetermined, it
 does not necessarily imply that there are no solutions for
 $K^{(2)}_{ijk}$, although we have not yet discovered any.

\subsection{Case iii)}

This is the case  of pure spin, as the momentum current vanishes.
Here the constraints are:                         \ba
  \psi_i & =& \pi_i   \approx 0\;,\cr
\phi_{ij}&=& \Sigma_{ij} + \frac12 L^{(3)}_{ijk\ell}
(\partial_\sigma \Lambda \Lambda^{-1})^{k\ell}
 \approx 0\;.\ea
From the second constraint and (\ref{sthre}),
$\Sigma_{ij}$ is once again identified with the time component of the
 spin current.   Since the orbital angular momentum vanishes,
$\Sigma_{ij}$ is  weakly equal to the total angular momentum
 ${\cal J}_{ij}$ defined in (\ref{jdef}).

The algebra of the constraints  now is given by:   \ba
\pois{\Psi[\lambda]}{\Psi[\lambda']}&=& 0\;,\cr
\pois{\Psi[\lambda]}{\Phi[\rho]}&=&0\;,\cr
\pois{\Phi[\rho]}{\Phi[\rho']}&\approx &
\int d\sigma \biggl(   ({\rho }^{ij} \partial_\sigma  {\rho'}^{k\ell}
-{\rho' }^{ij} \partial_\sigma  {\rho}^{k\ell})  L^{(3)}_{ijk\ell}
 -4\rho^{ij}{\rho '}_i^{\;k}  \Sigma_{jk}   \biggr)\;,\ea
 $\Psi[\lambda]$ and $\Phi[\rho]$  again being the
 distributions defined in (\ref{condis}).
   It immediately follows that    $\Psi[\lambda]$ for all $\lambda$
  are first class constraints.  In order to find the
remaining first class constraints we can once again define
     $\Theta [\hat \lambda,\hat\rho]$ as was done in (\ref{fcc}).  Then
 $\hat \rho$  satisfies:   \Eq{{{\hat \rho}_j} ^{\;k}
 \Sigma_{ik}    - {{\hat\rho}_i}^{\;k}\Sigma_{jk}-\frac12
  \partial_\sigma L^{(3)}_{ijk\ell} \hat\rho^{k\ell} \approx 0
   \;,\label{hat3}}
while $\hat\lambda$ is arbitrary.
Since  these equations are linear in  $\hat\rho$,
it  follows that if  $\hat\rho$ is   a solution, then so is
  $f\hat\rho$, where $f$ is an arbitrary function on the world sheet.
 Again the generators of diffeomorphisms of the world sheet can be
written in the general form:  $\Theta[f\hat \lambda,f\hat\rho]$,
here $\hat\lambda$ being arbitrary, while   $\hat\rho$ satisfies
 (\ref{hat3}).          A solution to the latter is
$\hat \rho=-\frac12 \partial_\sigma \Lambda \Lambda^{-1}$.  If we also
fix  $\hat \lambda$ to be equal to $ \partial_\sigma x$,
 we recover the generators of diffeomorphisms
${\cal L}[f] =\Theta[f\hat \lambda,f\hat\rho] $
on a fixed $\tau$ slice of the world sheet, which satisfy
 the Virasoro algebra (\ref{va}).  Thus we have again shown that
the Virasoro generators are first class constraints.

The canonical momenta $\pi_i$ have zero
Poisson brackets with all constraints,
and hence can be eliminated from the phase space.  The remaining
momenta $\Sigma_{ij}$ have the following Poisson brackets with
the constraints $\Phi[\rho]$:              \Eq{
 \pois{\Phi[\rho]}{\Sigma_{ij} (\sigma)}\approx - \partial_\sigma
(\rho^{k\ell}L^{(3)}_{ijk\ell}   )  \;.\label{sd3}}
We can choose  $H_\tau= \Phi[\hat\rho]$ for the Hamiltonian
 of this system, with    $\hat\rho$ being a solution to (\ref{hat3})
and the identification made of  $-\rho^{k\ell} L^{(3)}_{ijk\ell}   $
with the space component of the spin current $ S^1_{(3)ij}$ defined in
(\ref{sthre}).

From (\ref{sd3}), the angular momentum charges
\Eq{\int d\sigma\; \Sigma{ij}  }
have zero Poisson brackets with the constraints and hence are Dirac
variables.  Once again, they then generate the Lorentz algebra.
  We can construct the Dirac variables associated
with current densities $\Sigma_{ij}(\sigma) $,
in a similar manner to what was done previously.
 We  define
 \Eq{\tilde \Sigma[\xi] =  \int d \sigma\;\xi^{ij}(\sigma)
\Sigma_{ij}(\sigma) +\Phi[G_{(\xi)}]\;,} where
as before  $G_{(\xi)}$ denotes the functions
 $G_{(\xi)}^{ij}(\sigma)=-G_{(\xi)}^{ji}(\sigma)$.
For  $\tilde\Sigma[\xi]$
 to have zero Poisson brackets with all of the constraints, we need
that    $G_{(\xi)}$ satisfy:
\Eq{\partial_\sigma\xi^{k\ell} L_{ijk\ell}^{(3)} + G_{(\xi)}^{k\ell}
\partial_\sigma L^{(3)}_{ijk\ell}+2G_{(\xi)ik}{\Sigma_j}^k-
2G_{(\xi)jk}{\Sigma_i}^k\approx 0 \;.\label{Geq} }
Due to the existence of the first class constraints $\Phi[\hat\rho]$,
  $G_{(\xi)}$ are not uniquely defined by these equations.
Now the current algebra for the angular momenta has the following form:
\Eq{\pois {\tilde\Sigma[\xi]}   {\tilde\Sigma[\xi']} \approx
-4\tilde\Sigma [\xi\xi ']-4\tilde\Sigma [G_{(\xi)}G_{(\xi' )}]
+\int d \sigma\;  G_{(\xi )}^{ij} G_{(\xi ' )}^{k\ell}
  \partial_\sigma L^{(3)}_{ijk\ell}\;.\label{llg}}   This Poisson
bracket satisfies the antisymmetry property.

The first term on the right hand side of (\ref{llg})  corresponds
 to the algebra associated
 with the Lorentz loop group, while the remaining terms again
 represent an extension of the algebra.
    If we now demand that we should get
the Lorentz loop group algebra in the quantum
theory, then the above remaining terms are
 anomalous and they are obstructions
to the quantization of the theory.  Requiring that all the
anomalous terms vanish in (\ref{llg})  yields a total of
${1\over 8} d(d-1)(d+1)(d-2)$  conditions on the
${1\over 8} d(d-1)(d+1)(d-2)$
functions $L^{(3)}_{ijk\ell}$, and then on the
${1\over 8} d(d-1)(d+1)(d-2)$
constants $K^{(3)}_{ijk\ell}$.  The system is actually underdetermined
because, as stated earlier,
the functions   $G_{(\xi)}$ are not uniquely defined by
(\ref{Geq}).  We have not yet found solutions for
 $K^{(3)}_{ijk}$.

\subsection{Case iv)}
This is the most general case of a spinning string and it is simple to
put together all three of the previous cases.
The constraints  are now:                 \ba
\psi_i & =& \pi_i - L^{(1)}_{ij}\partial_\sigma x^j    -
 L^{(2)}_{ijk}(\partial_\sigma \Lambda \Lambda^{-1})^{jk}
  \approx 0\;,\cr
\phi_{ij}&=& \Sigma_{ij}    - 2L^{(2)}_{kij}\partial_\sigma x^k
 + \frac12 L^{(3)}_{ijk\ell}
(\partial_\sigma \Lambda \Lambda^{-1})^{k\ell}
 \approx 0\;,\ea  and their algebra is given by:
\ba\pois{\Psi[\lambda]}{\Psi[\lambda']}&=& \int d\sigma (\partial_\sigma
\lambda^j \lambda'^i -\partial_\sigma \lambda'^j \lambda^i) L^{(1)}_{ij}
\;,\cr
\pois{\Psi[\lambda]}{\Phi[\rho]}&\approx &
2\int d\sigma \rho^{ij} \biggl(
 \lambda_i \pi_j + \lambda^k(\partial_\sigma x_i L^{(1)}_{kj}
 - \partial_\sigma  L^{(2)}_{kij})\biggr) \; \;,\cr
\pois{\Phi[\rho]}{\Phi[\rho']}&\approx &  \int d\sigma \biggl(
-4{{\rho'}^i}_k \rho  ^{kj} \Sigma_{ ij}+4 (\rho^{ij}{\rho '}^{k\ell} -
{\rho '}^{ij}\rho  ^{k\ell})L^{(2)}_{\ell ij} \partial_\sigma x_k \cr
& & +   ({\rho }^{ij} \partial_\sigma  {\rho'}^{k\ell}
-{\rho' }^{ij} \partial_\sigma  {\rho}^{k\ell})  L^{(3)}_{ijk\ell}
 \biggr)\;,\ea

Once again we can find the subset of
 first class constraints for the system.
They have the general form:  $\Theta[f\hat \lambda,f\hat\rho]$.  If we
take $\hat \rho=-\frac12 \partial_\sigma \Lambda \Lambda^{-1}$ and
  $\hat \lambda= \partial_\sigma x$,
 we again recover the generators of diffeomorphisms
${\cal L}[f] =\Theta[f\hat \lambda,f\hat\rho] $
on a fixed $\tau$ slice of the world sheet, which satisfy
 the Virasoro algebra (\ref{va}).

The Poisson brackets of the momentum and angular momentum
current densities are again given by spatial derivatives.  We get
\ba\pois{\Psi[\lambda]}{\pi_i(\sigma)}&=& \partial_\sigma (\lambda^k
L^{(1)}_{ki}) \;,\cr
\pois{\Phi[\rho]} {\pi_i(\sigma)}&=&2\;
\partial_\sigma (L^{(2)}_{ijk}\rho^{jk})\;,\cr
  \pois{\Psi[\lambda]} {{\cal J}_{ij}(\sigma)  }
&\approx & \partial_\sigma\biggl(\lambda^k (x_i L^{(1)}_{kj} -
x_j L^{(1)}_{ki}
  -2 L^{(2)}_{kij})\biggr)  \;, \cr
\pois{\Phi[\rho]}{{\cal J}_{ij} (\sigma)}&\approx &-2 \; \partial_\sigma
\biggl(\rho^{k\ell}(L^{(2)}_{ik\ell} x_j - L^{(2)}_{jk\ell} x_i
         +\frac12
L^{(3)}_{ijk\ell} ) \biggr)    \;,\ea
from which it follows that the charges (\ref{char})
 generating the Poincar\'e
algebra are Dirac variables.
For the remaining Dirac variables we define
\Eq{\tilde \pi[\gamma]=\int d\sigma\; \gamma^i(\sigma)
\pi_i(\sigma)+\Psi[F_{(\gamma)}]+\Phi[E_{(\gamma)}]\;,}
where $E_{(\gamma)}^{ij}(\sigma)=-E_{(\gamma)}^{ji}(\sigma)$, along
with  $\tilde {\cal J}[\xi]$  in (66).
For  $\tilde \pi[\gamma]$ and $\tilde {\cal J}[\xi]$ to
have (weakly) zero Poisson brackets with  $\Psi[\lambda]$ and
 $\Phi[\rho]$, we
need that the functions $E_{(\gamma)}^{jk}$, $F_{(\gamma)}^k
$, $H_{(\xi)}^k$ and  $G_{(\xi)}^{jk}$  satisfy:
                                                      \ba
& &-\frac12 \partial_\sigma \gamma^i L^{(1)}_{ki}+ E_{(\gamma)kj}\pi^j
+E_{(\gamma)}^{ij}( L^{(1)}_{kj}  \partial_\sigma x_i  - \partial_\sigma
 L^{(2)}_{kij}) +\frac12  F_{(\gamma)}^j  \partial_\sigma L^{(1)}_{kj}
  \approx 0\;,\cr & &\cr
& &\partial_\sigma \gamma^i L^{(2)}_{ijk} +\frac12 (F_{(\gamma)j}\pi_k
-F_{(\gamma)k}\pi_j) -\frac12 F_{(\gamma)}^i(L^{(1)}_{ij}
\partial_\sigma x_k   -  L^{(1)}_{ik} \partial_\sigma x_j  + 2
\partial_\sigma   L^{(2)}_{ijk})\cr
& &+  E_{(\gamma)ji}{\Sigma_k}^i - E_{(\gamma)ki}{\Sigma_j}^i  -
E_{(\gamma)}^{\ell m}\biggl( L^{(2)}_{j\ell m}\partial_\sigma x_k  -
L^{(2)}_{k\ell m}\partial_\sigma x_j +2L^{(2)}_{ mjk}
\partial_\sigma x_\ell +\frac12 \partial_\sigma  L^{(3)}_{\ell mjk}
\biggr) \approx 0 \;,\cr & & \cr
& &\partial_\sigma\xi^{ij}( L_{kij}^{(2)} -x_i L_{kj}^{(1)})
-\frac12 H_{(\xi)}^i\partial_\sigma L^{(1)}_{ik}
 + G_{(\xi)ki}\pi^i +G_{(\xi)}^{ij} ( \partial_\sigma x_i L^{(1)}_{kj}
-\partial_\sigma L^{(2)}_{kij}) \approx 0\;,\cr & &\cr
& &\partial_\sigma\xi^{ij}\biggl(2L^{(2)}_{ik\ell}x_j
  +\frac12 L^{(3)}_{ijk\ell}\biggr)
-\frac12 (H_{(\xi)k}\pi_\ell  - H_{(\xi)\ell}\pi_k)
+\frac12 H_{(\xi)}^i(\partial_\sigma x_\ell L^{(1)}_{ik}
  - \partial_\sigma x_k L^{(1)}_{i\ell}+
 2 \partial_\sigma L^{(2)}_{ik\ell})\cr
& & -G_{(\xi)}^{ij}\biggl(L^{(2)}_{\ell ij}\partial_\sigma x_k
-L^{(2)}_{k ij}\partial_\sigma x_\ell -2
L_{jk\ell}^{(2)}\partial_\sigma x_i-\frac12 \partial_\sigma
L^{(3)}_{ijk\ell}\biggr)
-G_{(\xi)ik}{\Sigma^i}_\ell
+G_{(\xi)i\ell}{\Sigma^i}_k\approx 0 \;. \label{bgms}  \ea
Due to the existence of the first class constraints $\Theta$,
these functions are not uniquely defined by the above  equations.
The current algebra generated by the momentum
and angular momentum  densities is now given by:    \ba
\pois{\tilde \pi[\gamma]}{\tilde \pi[\gamma']} &\approx &
\int d\sigma\; \partial_\sigma \gamma'^i( L^{(1)}_{ij}F_{(\gamma)}^j
-2  L^{(2)}_{ijk}   E_{(\gamma)}^{jk} )\;,\cr
\pois {\tilde {\cal J}[\xi]} {\tilde \pi[\gamma]}
&\approx & -2\tilde\pi [\xi \gamma] +
\int d\sigma\; \partial_\sigma \gamma^i( L^{(1)}_{ij}H_{(\xi)}^j
-2  L^{(2)}_{ijk}   G_{(\xi)}^{jk}) \;,\cr
\pois {\tilde {\cal J}[\xi]}   {\tilde {\cal J}[\xi']} &\approx  &
-4\tilde {\cal J} [\xi\xi '] -2
\int d\sigma \; \partial_\sigma \xi'^{ij}\biggr((x_i L^{(1)}_{kj}
- L^{(2)}_{kij}) H_{(\xi)}^k
+(2x_i L^{(1)}_{jl\ell}   -\frac12  L^{(3)}_{ijk\ell})  G_{(\xi)}^{k\ell}
\biggr)\;.\cr & &\ea  The  $\tilde \pi-\tilde \pi $ and
$\tilde {\cal J}-\tilde {\cal J}$
Poisson brackets  can be shown to satisfy the antisymmetry
property using (\ref{bgms}).

Once again we get a nontrivial extension  Poincar\'e loop group algebra.
Now requiring that all the
anomalous terms vanish   yields a total of
${1\over 8} d(d-1)(d+1)(d+2)$  conditions on the
${1\over 8} d(d-1)(d+1)(d+2)$
functions $L^{(1)}_{ij}$, $L^{(2)}_{ijk}$, $L^{(3)}_{ijk\ell}$,
 and then on the ${1\over 8} d(d-1)(d+1)(d+2)$  constants
$K^{(1)}_{ij}$, $K^{(2)}_{ijk}$, $K^{(3)}_{ijk\ell}$.
As in the previous case, the system is actually underdetermined
because the functions  $E_{(\gamma)}^{jk}$, $F_{(\gamma)}^k
$, $H_{(\xi)}^k$ and  $G_{(\xi)}^{jk}$
 are not uniquely defined by (\ref{bgms}).  Again,  we have not yet
 found solutions for   the constants $K$.

\section{Generalizations to Membranes}

Here we show how  to generalize the action for spinning strings to
 higher dimensional spinning objects, like membranes.  The system
again contains interesting special cases such as the spinless membrane,
where the action is equivalent to the world volume, as well as the case
of pure spin, where a spin current is present but the
 momentum current vanishes.  We do not analyze the Hamiltonian dynamics
for such systems here.

If, like the case with spinning particles and spinning strings,
 we let the action  depend only on
 the components $V$ and $W$ of Maurer-Cartan form $g^{-1}dg$,
 it will automatically
 be invariant under (left) Poincar\'e transformations (\ref{lpt}).
Geometric actions for  $p-$dimensional objects are obtained by
constructing $p-$forms from
  $V$ and $W$ defined in eq.
  (\ref{defVW}).  There are now $p+1$ possible terms:
\Eq{S=S_{K}(\Lambda, x)=S_1+S_2+...+S_{p+1}  \;,} where
\ba S_1&=&\int K^{(1)}_{i_1i_2...i_p}\; V^{i_1} \wedge V^{i_2}
\wedge...\wedge V^{i_p}\;, \label{slcfm}  \\
S_2&=&\int K^{(2)}_{i_1j_1i_2i_3...i_p}\; W^{i_1j_1} \wedge V^{i_2}
\wedge...\wedge V^{i_p}\;,\\     & &............\cr   & & \cr
S_{p+1}&=& \int K^{(p+1)}_{i_1j_1i_2j_2...i_pj_p}\; W^{i_1j_1}
\wedge  W^{i_2j_2}\wedge ...\wedge  W^{i_pj_p}\;,\ea
where $K=(K^{(1)}_{i_1i_2...i_p},\;K^{(2)}_{i_1j_1i_2i_3...i_p},...,\;
 K^{(p+1)}_{i_1j_1i_2j_2...i_pj_p})$ denotes a set of
constants.

Like with the case of strings, the action $S_1$  by itself
 describes a spinless
object, and was discussed in \cite{baletal}.  This  result  is
easily checked from the corresponding equations of motion.
On the other hand,
all the remaining terms give rise to a nonvanishing spin current.  The
action  $S_{p+1}$ by itself has only the spin current contributing
to the total angular momentum, as its associated
 momentum current vanishes.

\bigskip
\bigskip

{\parindent 0cm{\bf Acknowledgements:}}
A. S. is grateful to R. Casadio for useful  discussions,
 and was supported in part
by the Department of Energy, USA, under contract number
DE-FG05-84ER40141.

\end{document}